# Superstructure evolution of tellurium atoms on Au(111) surface at different coverages

Jiaqi Guan[1,2], Xiaochun Huang[1,2], Shuyuan Zhang[1,2], Xun Jia[1,2], Xuetao Zhu[1], Weihua Wang[1*] and Jiandong Guo[1,2,3*]

[1]*Beijing National Laboratory for Condensed Matter Physics and Institute of Physics, Chinese Academy of Sciences, Beijing 100190, China*

[2]*School of Physical Sciences, University of Chinese Academy of Sciences, Beijing 100190, China*

[3]*Collaborative Innovation Center of Quantum Matter, Beijing 100871, China*

*Email: weihuawang@iphy.ac.cn and jdguo@ iphy.ac.cn

**Abstract**

We systematically investigated the superstructure evolution of Te atoms on Au(111) substrate at different coverages. As revealed by low temperature scanning tunneling microscopy and spectroscopy, Te atoms form one-dimensional $\sqrt{3}R30°$ chains near 0.10 monolayer (ML). Two two-dimensional chiral structures, $(\sqrt{111} \times \sqrt{111})R4.7°$ and $(3\sqrt{21} \times 3\sqrt{21})R10.9°$, can be formed and their stability can be tuned by slightly adjusting the Te coverge near 1/3 ML. A honeycomb-like superstructure is observed by further increasing the coverage to 4/9 ML. An interfacial state emerges at ~-0.65 eV due to Te adsorption on Au(111). The formation of these Te-induced high-order superstructures is accompanied by relaxation of gold atoms in the surface layer, indicating the strong Te-Au interaction.

# 1. Introduction

Gold surface is widely used as a template in atom and molecule adsorption [1-3], self-assembly [4, 5], surface-supported reactions [6, 7], and also believed to be applicable in chemical catalysis and energy storage [8, 9]. On the reconstructed Au(111) surface with herringbone pattern, a consequence of the formation of mesoscopic stress domain on the highly close-packed surface [10, 11], the adsorption exhibits intriguing physical insights as the result of adsorbate-substrate interaction. The surface reconstruction of Au(111) can be influenced by adsorption. More specifically, previous reports have shown that oxygen, sulfur, chlorine, tellurium, gadolinium atoms and trimethylphosphines molecules can compensate the residual tensile strain of the clean surface by introducing compressive stress [12-19], while alkali metal atoms can enhance the corrugation of topmost Au layer [20]. Additionally, the surface states of Au(111) can be modified by adsorbates [21-24], offering the tunability of properties of the Au template.

Tellurium is the heaviest non-radioactive element in chalcogen group, which introduces strong spin-orbit coupling to various tellurides [25-31]. The bulk Te crystal is a mid-infrared semiconductor with an energy gap of 0.33 eV [32], and theoretical works predict its topological properties under strain or pressure [33, 34]. Recently, telluride film has been synthesized by depositing tellurium atoms on hafnium, opening up a new route to the fabrication of functional tellurides nanostructures and a platform for the futher study of novel quantum phenomena [35]. The adsorption of Te atoms on gold has been investigated by various experimental methods, including low-energy electron diffraction (LEED), electrochemical scanning tunneling microscopy (EC STM), auger electron spectroscopy (AES) and atomic force microscopy (AFM) [36-42]. Previous LEED study indicates that the Te layer on Au(111) exhibits a $\sqrt{3} \times \sqrt{3}$ reconstruction [17]. The STM and AFM studies revealed several high-order superstructures depending on the Te coverage [17, 38-40]. Most recently, K. Schouteden and co-workers studied the adsorption of Te atoms on Au(111) at extremely low coverage, and revealed the strong interaction between Te adatoms and Au(111) surface, leading to scattering and confinement of the Au surface state near the Fermi level [43]. However, a systematic study of Te adsorption on Au(111) surface with increased coverage is still lack.

Here, by precisely controlling the Te coverage, we systematically investigated the superstructure evolution of Te atoms on Au(111) substrate by *in-situ* low-temperature STM and scanning tunneling spectroscopy (STS). Te atoms form one-dimensional (1D) chains at low coverage, and two-dimensional (2D) superstructures at higher coverage, including two chiral structures near 1/3 monolayer (ML) and a honeycomb-like structure at ~4/9 ML. The formation of these superstructures are accompanied by the relaxation of Au(111) reconstructed lattice. Moreover, we detect the interfacial states of the Te/Au(111) system, which manifests the interaction between Te and Au surface.

## 2. Experiments

Our experiments were performed in an ultrahigh vacuum low-temperature scanning tunneling microscope with a base pressure better than $1\times10^{-10}$ Torr. The Au(111) substrate was cleaned by cycles of $Ar^+$ sputtering and annealing, which led to $(22\times\sqrt{3})$ reconstruction as verified by STM. Te (99.999+%, Alfa Aesar) was evaporated from a Knudsen cell (CreaTec) at 500 K, and deposited on Au(111) substrate held at room temperature. The evaporation rate was calibrated by STM as 0.083 ML/minute [1 ML is defined as the atom amount on the bulk-truncated Au(111) surface], since each Te adatom appears as a bright protrusion in the image [43]. After Te deposition, the sample was transferred into the STM and cooled down to 4.9 K. Polycrystalline Pt-Ir tips were used throughout the experiments. STM topographic images were acquired in constant-current mode, and differential conductance (d$I$/d$V$) signals were acquired with a sinusoidal modulation of 10 meV at 987.5 Hz.

## 3. Results and discussion
### 3.1 One-dimensional $\sqrt{3}R30°$ chains

With normal preparation procedure, a clean Au(111) surface shows the $(22\times\sqrt{3})$ reconstruction, or herringbone reconstruction, in which every 23 Au atoms are squeezed into 22 bulk lattice sites on the surface [10, 44]. In order to release the surface stress, Au atoms occupy fcc, fcp and bridge sites to increase the surface area, resulting in residual tensile stress on the reconstructed surface [11, 12]. The domain walls, also known as soliton walls, which separate the wide fcc and the narrow hcp regions, have an increased density of Au atoms and appear as bright stripes in STM image, as shown in Fig. 1(a). These domain walls extend along the $\langle 11\bar{2}\rangle$ directions, with a well-defined separation of 6.34 nm between every pair of domain walls along the $\langle 1\bar{1}0\rangle$ directions. Some localized structures on the $(22\times\sqrt{3})$, e.g., partial dislocations at elbows of the soliton walls, can further minimize surface free energy [10, 11].

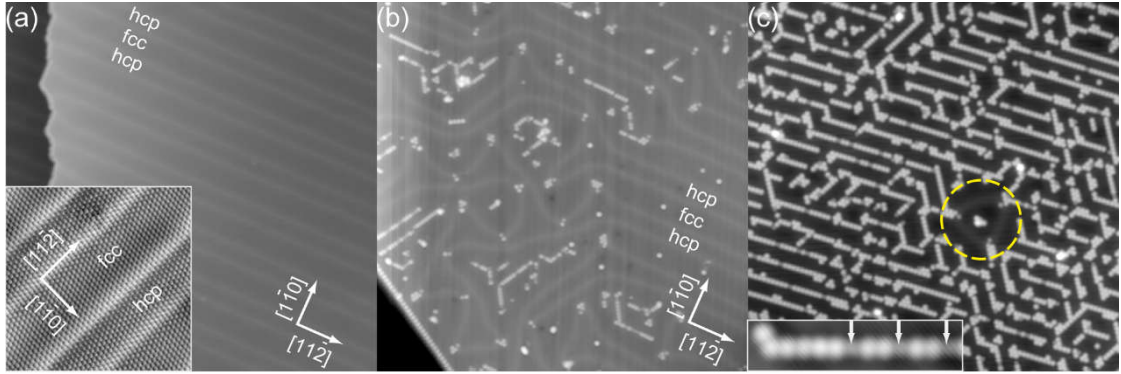

Figure 1 (Color online) (a) STM image (50×50 nm², 3.0 V/50 pA) of clean Au(111). The inset shows the atom-resolved image (1.0 V/200 pA). (b) and (c) STM images of Te atoms on Au(111) at coverage of 0.02 ML (50×50 nm², 1.2 V/70 pA) and 0.10 ML (40×40 nm², 1.0 V/50 pA), respectively. The inset shows the magnified STM image (6×1.4 nm², 0.2 V/50 pA) of Te chain on Au(111) with white arrows indicating the dim Te atoms.

At a low coverage of 0.02 ML, Te atoms tend to occupy elbows and fcc domains on the surface (besides step edges), as seen in Fig. 1(b). Te atoms appear as isolated atoms or short chains along the $\langle 11\bar{2}\rangle$ directions. The Au(111) substrate shows disturbed herringbone structures in contrast to the ordered soliton walls before Te deposition [Fig. 1(a)], and the fcc regions with Te chains are broadened. These results are in agreement with a very recent work with residual Te atoms on Au(111) surface [43]. With Te coverage increased to 0.10 ML, Te atomic chains extended longer along the $\langle 11\bar{2}\rangle$ directions, and the space between two neighboring Te atoms is ~0.5 nm, corresponding to $\sqrt{3}R30°$ chains along the $[11\bar{2}]$ direction.

Previous reports have shown that electronegative adsorbates such as sulfur, oxygen, and chlorine compensate the residual tensile stress of Au(111) surface and disturb or even suppress the reconstruction [12-14], while alkali metal atoms enhance the corrugation of topmost Au layer [20]. In the current work of the adsorption of electronegative Te atoms, the residual strain on reconstructed Au(111) surface is expected to be relieved by the interaction between Te and the substrate. As a result, we observe that the soliton walls are almost completely removed from the surface, with only small triangle-shaped areas left, as highlighted by the dashed circle in Fig. 1(c).

**3.2 Chiral structures and interfacial state**

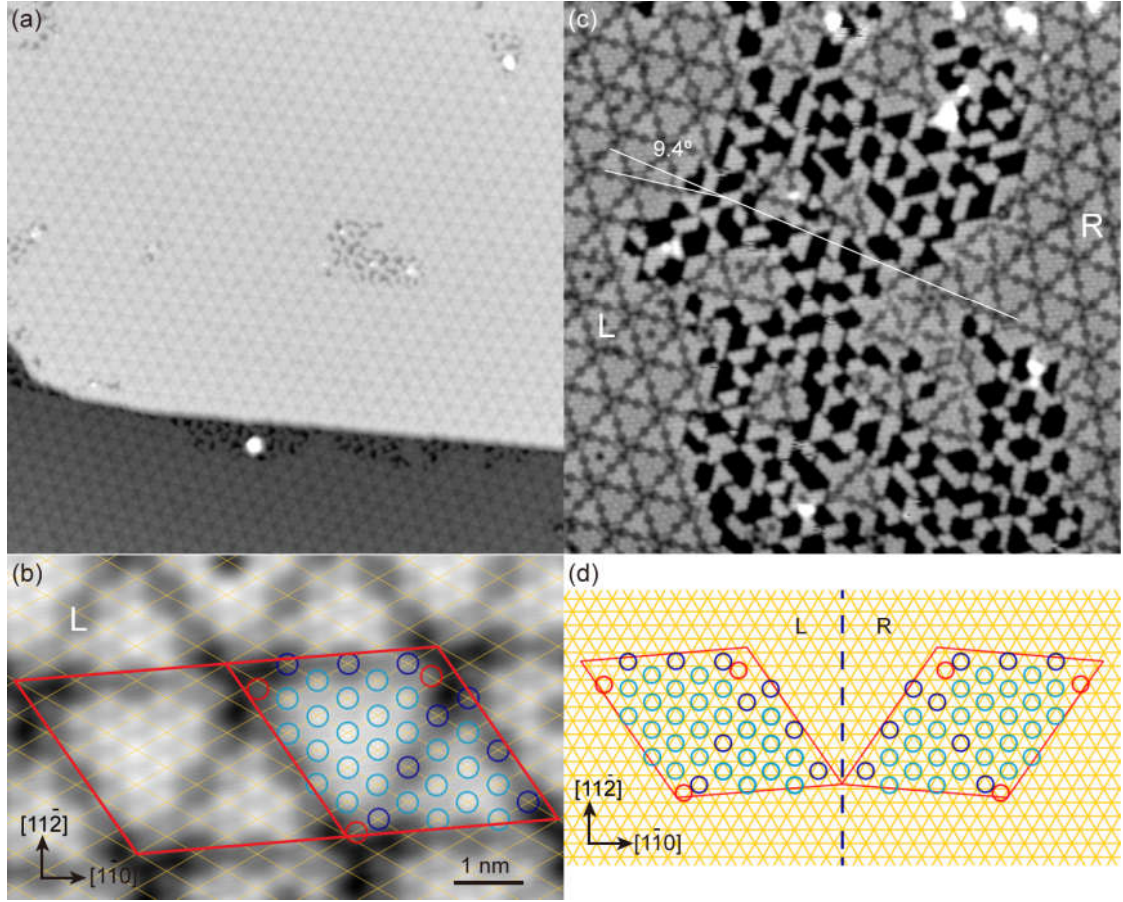

Figure 2 (Color online) (a) STM image (100×100 nm$^2$, 2.0 V/50 pA) of Te atoms on Au(111) at coverage of 0.24 ML. (b) Zoom-in STM image (8×5 nm$^2$, -0.025 V/50 pA) of the rhombus phase. The $\sqrt{3} \times \sqrt{3}$ lattice of Au(111) are indicated by the yellow grids, and the $(\sqrt{111} \times \sqrt{111})R4.7°$ unit cells are shown by the red frames. (c) STM image (40×40 nm$^2$, 1.5 V/100 pA) showing two types of domains coexist with each other. The left and right chiralities are illustrated by the yellow lines. (d) Adsorption model of the $(\sqrt{111} \times \sqrt{111})R4.7°$ rhombus phase. The bright Te atoms are represented by cyan circles, the dim Te atoms by violet and red circles. The cyan and violet circles are sitting on the same set of three-fold hollow sites, while the red circles are sitting on the other set.

When Te coverage increased to 0.24 ML, the Au terrace is covered by a 2D rhombus phase, as shown in Fig. 2(a). Figure 2(b) displays a zoom-in STM image, in which the Te atoms are clearly resolved. A $\sqrt{3} \times \sqrt{3}$ lattice of Au(111) substrate is superposed on the image. Most of Te atoms are sitting on the $\sqrt{3} \times \sqrt{3}$ lattice sites, except for those indicated by red circles. The rhombus phase corresponds to a $(\sqrt{111} \times \sqrt{111})R4.7°$ (*root*-111) superstructure, which has a periodicity of 3.04 nm as indicated by red frames. Each rhombus unit cell contains two Te quasitriangles formed by bright Te atoms [indicated by cyan circles in Fig. 2(b)] and dim atoms between them [indicated by violet and red circles]. The bright and dim features are irrespective of scanning conditions. The adsorption model is illustrated in Fig. 2(d). Each unit cell contains 37 Te atoms, 34 of which are adsorbed on the same set of $\sqrt{3} \times \sqrt{3}$ lattice, *i.e.*, the fcc (or hcp) three-fold hollow sites, while other 3 atoms in each unit cell are adsorbed at the other set of hcp (or fcc) three-fold hollow sites.

Two types of the *root*-111 superstructure domains coexist with each other on the surface, as shown in Fig. 2(c). The two domainshave an angle of 9.4° (indicated by the white lines). A detailed inspection of the image shows that the two domians have different chiralities, as illustrated by yellow lines in Fig. 2(c). The domain L can be defined as they are arranged clockwisely (left part), while domain R is defined as they are arranged anti-clockwisely (right part). Since the L and R domains are rotated from the $[1\bar{1}0]$ direction by ± 4.7°, the two domains has an angle of 9.4°. And the *root*-111 superstructure is always accompanied by poroused areas at the domain boundary [see Fig. 2(c) and (d)].

When Te coverage slightly exceeds 1/3 ML, a new 2D rhombus phase with longer periodicity is formed on the surface. As shown in Fig. 3(a), this new phase is accompanied by interstitial Te atom or clusters in the holes at rhombus corners. This rhombus phase also has chirality, as illustrated by yellow lines in Fig. 3(a) and 3(b), corresponding to the L and R domains, respectively.

Figure 3(a) and 3(b) shows that bright and dim Te atoms coexisting with each other, with dim atoms distributed at the rim of quasitriangles formed by bright Te atoms. In contrast to the *root*-111 phase, all the bright and dim Te atoms in this phase are sitting on the same set of $\sqrt{3} \times \sqrt{3}$ lattice. Illustrated in the zoom-in STM image of Fig. 3(b), this new rhombus phase is represented by a $(3\sqrt{21} \times 3\sqrt{21})R10.9°$ (*root*-189) superstructure with a periodicity of 3.96 nm. The proposed adsorption model is depicted in Fig. 3(c), and each unit cell contains 62 Te atoms if the holes at the corners are not occupied by Te atoms or clusters.

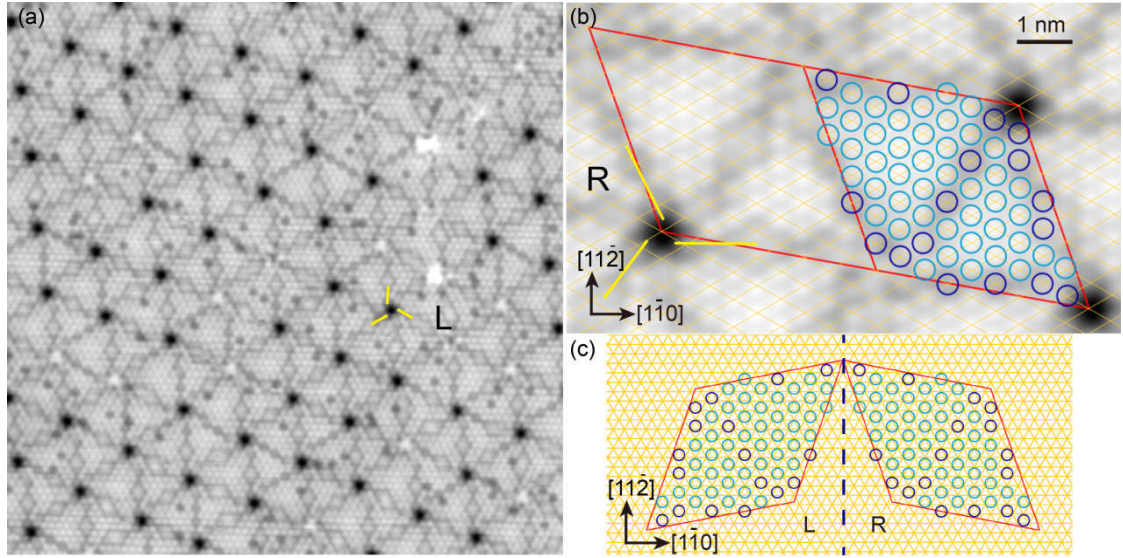

Figure 3 (Color online) (a) STM image (40×40 nm$^2$, 1.0 V/50 Pa) of the other rhombus phase with enlarged unit cell. (b) Zoom-in STM image (10×6 nm$^2$, 1.0 V/100 pA). The $\sqrt{3} \times \sqrt{3}$ lattice of Au(111) substrate are indicated by yellow grids, and the unit cells of the rhombus phase are shown by red frames. The chiralities of the rhombus phases in (a) and (b) are illustrated by yellow lines. (c) Adsorption model of the rhombus phase. The bright (dim) Te atoms in STM images are represented by cyan (violet) circles in (b) and (c).

Experimentally, the *root*-111 and *root*-189 phases can be converted to each other reversibly. The *root*-111 phase can be converted to *root*-189 by adding Te atoms, and changed back by moderate annealing at about 200°C to reduce the Te coverage.

It is interesting to compare the nominal Te coverage of the two rhombus phases. The *root*-111 contains 37 Te atoms per 111 Au atom sites, giving a nominal Te coverage of 1/3 ML, while the *root*-189 phase has a nominal Te coverage of 0.328 ML. If each unit cell of *root*-189 phase hold a Te atom in its corner, the nominal Te coverage is 1/3 ML, identical to the *root*-111 phase. In reality, the *root*-111 phase is accompanied by poroused areas that lower the Te density significantly. As the result, the Te coverage of *root*-111 phase is below 1/3 ML. In contrast, since *root*-189 phase can hold Te atoms or even clusters at the corners of each rhombus unit cell, this phase is stable with Te coverage a little higher than 1/3 ML.

It is noted in both rhombus phases dim Te atoms are observed at the rim of quasitriangles formed by bright Te atoms. This is related to the lattice compression of the topmost Au layer induced by Te atoms. In the 1D case when the Te coverage is low [~0.10 ML, see Fig. 1(c)], the Te chains modify the Au-Au distance in 1D, leading to the existence of enlarged inter-atom distance at certain sites. As the result, the Te atoms on these sites will be lower than others, appearing as dim atoms in STM [inset of Fig. 1(c)]. With increased Te coverage, Te atoms form 2D superstructures. Accordingly the sites with enlarged Au-Au distance will be distributed in 2D, resulting in the observed distribution of dim Te in STM images [Fig. 2(c) and Fig. 3(b)]. Similar mechanism has been reported on Cl-adsorbed Au(111) surface, on which a quasihexagonal superstructure was formed [14].

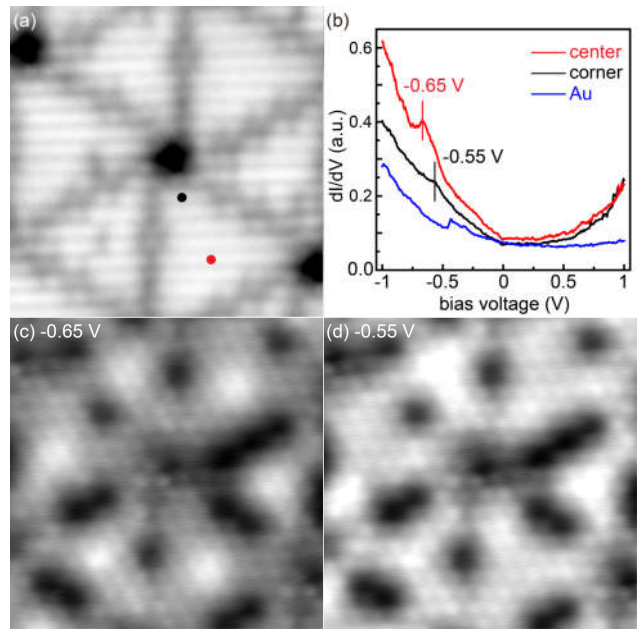

Figure 4 (Color online) (a) STM image (8×8 nm$^2$, -0.65 V/500 pA) of *root*-189 phase. (b) Averaged STS spectra measured at the center and corner positions, as indicated by the red and black dots in (a), respectively. A spectrum measured on clean Au(111) surface is shown for comparison. (c) and (d) d$I$/d$V$ maps of the same area in (a) measured at -0.65 V and -0.55 V, respectively.

The Te-Au interaction gives rise to not only the surface relaxation, but also the formation of interfacial states. As shown in Fig. 4(a) and 4(b), the d$I$/d$V$ spectrum measured at the center of the triangle in *root*-189 phase shows a new electronic state at -0.65 V induced by Te adsorption, which shifts to -0.55 V at the corner, both at energies lower than the surface state onset of claen Au(111) surface [-0.44 V, see blue line in Fig. 4(b)]. The distribution of this state in real space is clearly resolved by the d$I$/d$V$ maps at -0.65 V and -0.55 V, as shown in Fig. 4(c) and 4(d), respectively, indicating a dispersive electronic state confined in the Te triangle.

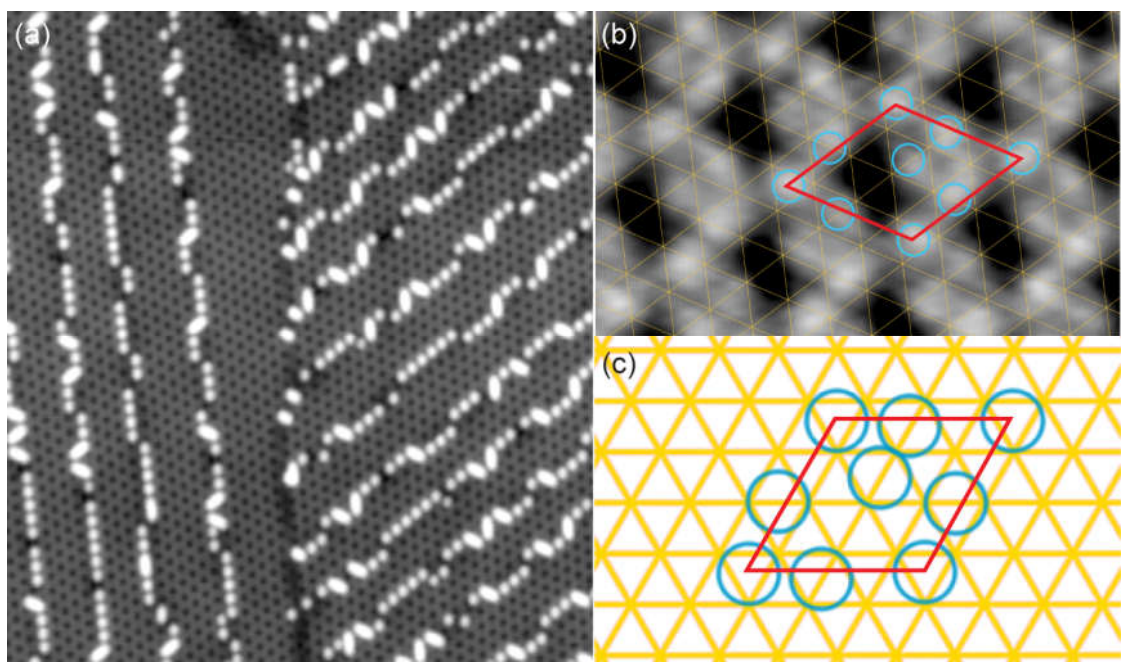

Figure 5 (Color online) (a) The STM image (40 × 40 nm$^2$, 1.0 V/80 pA) of the honeycomb-like phase with increased Te coverage. (b) Atomically resolved images [4×2.5 nm$^2$, 200 pA/0.01 V] of the honeycomb structure. (c) Adsorption model of the honeycomb-like phase, with the red frame indicating the unit cell.

### 3.3 Honeycomb structure

When the Te coverage is further increased, a honeycomb-like phase is formed on the substrate, with adatoms distributed in quasi-1D chains, as shown in Fig. 5. The inter-chain spacing is relatively uniform, ranging from 3 nm to 10 nm. The atomic structure of the honeycomb-like phase is resolved in the STM image shown in Fig. 5(b). By superposing the Au(111) lattice on the images, the honeycomb-like phase is found to be a (3×3) superstructure. The adsorption model is illustrated in Fig. 5(c). The Te atoms occupy not only the hollow sites but also the bridge sites. Each (3×3) unit cell contains 4 Te atoms, giving a nominal Te coverage of 4/9 ML.

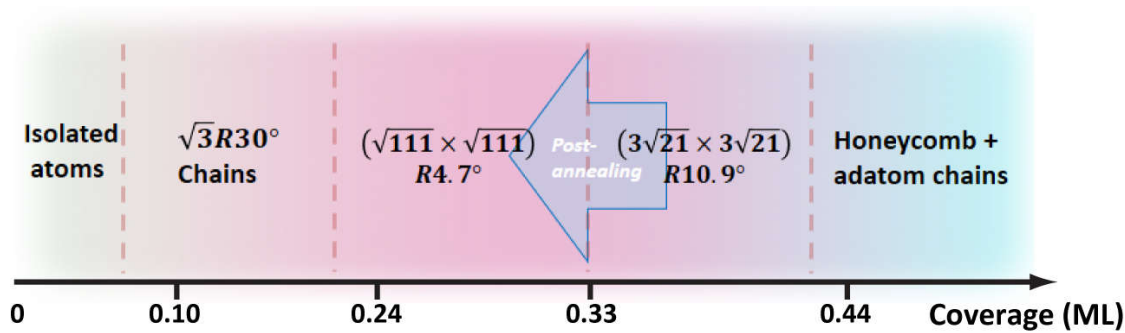

Figure 6 (Color online) Phase diagram of Te adsorption structures on Au(111) at different coverages.

## 4. Conclusions

The phase diagram shown in Fig. 6 illustrates the morphological evolution of Te on Au(111) at different coverages, from 1D chains to 2D superstructures. The formation of these ordered phases is dictated by the strain relief mechanism of Au(111) surface. Compared with the clean Au(111) surface, Te-adsorbed Au(111) exhibits an interfacial state at ~-0.65 eV. With precisely controlled Te coverage on Au(111), the adsorbate-substrate interaction can be directly probed. Our work demonstrates a reliable method to fabricate Te nanostructures on noble metal surface in a controlled way.

Weihua Wang thank Prof. Banggui Liu for fruitful discussions. This work is supported by the National Key Research and Development Program of China (2016YFA0300600, 2016YFA0202300), Chinese NSFC (11634016) and the Hundred Talents Program of the Chinese Academy of Sciences.